%% file: arduinopumppaper.tex
\documentclass[review]{elsarticle} 
\usepackage[english]{babel}
\usepackage{graphicx}
\usepackage{pstool}
\usepackage{eurosym}
\usepackage{lineno}

\begin{document}
\title{Arduino control of a pulsatile flow rig}
\author[ulb]{S.\ Drost}

\author[uld]{B.\ J.\ de Kruif}

\author[ulb]{D.\ Newport\corref{cor1}}
\ead{david.newport@ul.ie}

\cortext[cor1]{Corresponding author}

\address[ulb]{Bernal Institute, School of Engineering, University of Limerick, Limerick, Ireland}
\address[uld]{Design Factors, Faculty of Science and Engineering, University of Limerick, Limerick, Ireland}

\begin{abstract}
This note describes the design and testing of a programmable pulsatile flow pump using an Arduino micro-controller. The goal of this work is to build a compact and affordable system that can relatively easily be programmed to generate physiological waveforms. The system described here was designed to be used in an in-vitro set-up for vascular access hemodynamics research, and hence incorporates a gear pump that can deliver up to around 1.5~l/min (steady flow) in a test flow loop. After a number of simple identification experiments to assess the dynamic behaviour of the system, a feed-forward control routine was implemented. The resulting system was shown to be able to produce the targeted patient-specific waveform with less than 3.6\% error. Finally, we outline how to further increase the accuracy of the system, and how to adapt it to specific user needs.
\end{abstract}


\maketitle

%\linenumbers

\input{introduction.tex} 
\input{method.tex}			
\input{experiments.tex}	
\input{conclusions.tex}

\section*{Acknowledgments}
This project has received funding from the European Union’s Seventh Framework Programme for research, technological development and demonstration under grant agreement no. 324487.

Competing interests: none declared.

Ethical approval: not required.

\nolinenumbers

\bibliographystyle{plain}
\bibliography{refs}
\end{document}

%% file: introduction.tex
\section{Introduction}
\label{sec:Introduction}
In-vitro experiments are a useful tool in hemodynamics research, offering a wide range of applicable experimental techniques, combined with good accuracy and reproducibility. To simulate physiologically realistic pulsatile flow, a dedicated pumping system is used in these experiments. This can either be a mock loop representing the full circulatory system \cite{ref:Schampaert14,ref:Segers98}, or a pump that reproduces a waveform measured locally in-vivo \cite{ref:Chaudhury16,ref:Hoskins89,ref:Petersen84,ref:Plewes95,ref:Tsai10}. 

For both cases, pumps are commercially available. However, researchers often employ in-house built systems, for example because of cost considerations, or to add specific functionality. Examples of pumping systems that were developed to reproduce physiological waveforms are a gear pump driven by a stepper motor or servomotor, the rotation rate of which is controlled by a microcomputer (Hoskins et al. \cite{ref:Hoskins89}) or personal computer (Plewes et al. \cite{ref:Plewes95}), a controlled piston pump (Chaudhury et al. \cite{ref:Chaudhury16}), or a combination of both: a gear pump to deliver the steady flow component, and a piston pump to generate the pulsatile component (Tsai and Sava\c{s} \cite{ref:Tsai10}).

With so many options already available, the goal of our current work is not so much to develop a superior system as to design a system that is affordable and relatively simple to set up, using off-the-shelf components. With this in mind, the full source code for the control of the pump is included as a supplement to this note. 

The system that is presented in this technical note is designed to reproduce a waveform typical for an arterio-venous fistula for hemodynamics, more specifically, a waveform with a mean flow of $\mathcal{O}(1)$~l/min, a frequency content up to 4~Hz, and a pulsatility index (peak-to-peak flow over mean flow) of 0.4. We consciously chose to use the simplest possible control strategy, first identifying the system, and subsequently using a feed-forward controller in combination with a run-to-run controller to achieve the target waveform. The present system is aimed to be used in an in-vitro flow rig to test different modalities for vascular access for hemodynamics, but with suitable parts selection, the approach we used is expected to be useful for a wider range of cardiovascular flows as well.

%% file: method.tex
\section{Materials and method}
\subsection{Set-up}
\begin{figure}
\centering
	\includegraphics[width=0.5\textwidth]{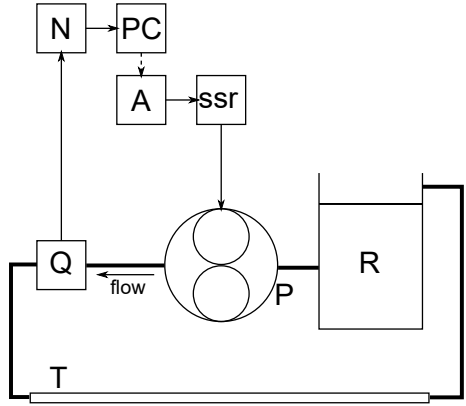}
	\caption{Experimental set-up (not to scale): Flow loop, consisting of reservoir (R), flexible tubing, perspex tube (T), Transonic flow sensor (Q), and gear pump (P). Control loop, consisting of an  Arduino (A) with a Solid State Relay (SSR). For monitoring and identification a data acquisition card (N) and PC are used.}
	\label{fig:pump_schematic}
\end{figure}

The experimental set-up is shown schematically in Figure~\ref{fig:pump_schematic}. The flow loop that was used for the experiments consists of a straight poly(methyl methacrylate) (PMMA) tube with an inner diameter, ID, of 5~mm and a length of 500~mm, connected to an open reservoir ($\pm 500$~ml) and to the pump by flexible polyethylene tubing (ID 6~mm). This is the simplest possible set-up to study fully developed steady or pulsatile laminar flow, up to a Reynolds number of 2000, while the ID of the straight tube is representative for vascular access. Tap water at room temperature was used as a test liquid. An in-line flow sensor (Transonic TS410 ME 6PXN), placed around 50~cm from the pump outlet, was used to monitor the flow rate, its signal was sent to a data acquisition card (National Instruments, NI USB-6001), which in turn was read out using Matlab. Even though, when used in the prescribed range of rotation rates, gear pumps produce negligible ripple, care was taken to place the flow sensor sufficiently far from the pump outlet.

The pumping system consists of a gear pump (excess from the medical manufacturing and spare parts for a Gambro C3/CS3 dialysis machine, equivalent to a Diener Extreme Series 1000~ml/min with 9~mm gear set), magnetically driven by a core-less DC motor (Premotec, 30~V DC). A 24~V DC power supply is used, in combination with a solid-state relay (SSR) to control the voltage supplied to the motor. This setup results in a one quadrant controller that can only apply a positive torque with a positive velocity. 

An Arduino micro-controller (Genuino MEGA 2560) sends a pulse width modulation (PWM) signal to the SSR to achieve the required waveform during the experiment. The Arduino has a resolution of 8 bits, which means that the PWM signal can be set to integer values between 0 and 255, to achieve a motor voltage between 0 and 24~V. To be more flexible during the design phase of the system, we used a PC with Matlab to send the calculated pump voltages to the Arduino. To emphasise that this communication only happens during the design phase, the connection in Figure~\ref{fig:pump_schematic} is dashed. Arduino offers the option of a Secure Digital (SD) card module, so that eventually the waveform can be stored on an SD-card and the Arduino can operate as a stand-alone controller.

The total cost of the system (excluding the Transonic system) was approximately \euro~350 ($\approx$ \$ 393: $\pm$ \euro 250 for the pump and motor, \euro 50 for the Arduino, and \euro 50 for the remaining parts), which is considerably less than the typical cost for a commercial system ($\pm$ \$ $\mathcal{O}(10^4)$, estimate based on personal communication with suppliers).

\subsection{Method}
The targeted velocity waveform for our experiments is representative for the waveforms encountered at the proximal arterial side of an arterio-venous fistula used for hemodialysis \cite{ref:BrowneThesis,ref:Browne15,ref:Sigovan13,ref:Sivanesan99}. It is constructed as a Fourier series prescribing the flow rate, $Q(t)$, in l/min: 
\begin{equation}
Q(t) = \sum_{n = -N}^N c_n \exp\left(i\frac{2\pi n t}{T}\right),
\end{equation}
with a period, $T$, of 1~s, $N = 4$, and coefficients, $c_n$ given in table \ref{tab:coeff} (for a graphical representation see Figure \ref{fig:MeasRefFilt}).
\begin{table}
\centering
\centering
\caption{Coefficients for Fourier series to construct waveform ($c_n^*$ denotes the complex conjugate of $c_n$).}
\label{tab:coeff}
	\begin{tabular}{l l}
	$n$ & $c_n$ \\
	\hline
	0 & 0.9 \\
	1 & $-0.015 - $i \\
	2 & $-0.036 - 0.03$i \\
	3 & $-6\cdot 10^{-3} + 1.2\cdot 10^{-3}$i \\
	4 & $-0.012 + 6\cdot 10^{-4}$i \\
	$c_{-n}$ & $c_{n}^*$ \\
	\end{tabular}
\end{table}
To be able to reproduce this waveform, the non-linear and dynamic behaviour of our system are identified, and a controller is designed based on this.

\subsubsection{System identification}
The system is considered as a combination of a linear dynamic system with a static non-linear output function. The equations that describe this behaviour are:
\begin{equation}
	\tilde{Q} = P(s)u, \ Q = f(\tilde{Q})\label{eq:model}.
\end{equation}
In this, $\tilde{Q}$ denotes a virtual flow (in arbitrary units), while $Q$ represents the true flow in litres per minute, $s$ is the Laplacian variable, and $P$ the linear transfer function of the system. The input $u$ is the voltage to the pump. Although a first principles physical model is not made of the system, the non-linearitiy expected to be dominant is the flow - gear wheel relation due to back flow (``slip'': not all fluid on the intake side is transported to the outlet side by the gears. This effect is particularly noticeable at higher rotation rates, and for low viscosity working fluids).

In order to identify the relation in (\ref{eq:model}) two dedicated experiments are performed. The output non-linearity is identified by applying a series of different input voltages to the pump and measuring its steady state output. A smooth function is fitted through these measurement points, minimising the weighted sum-of-squares error. The weighting in this fitting procedure is done to optimise the fit in the range of flow rates present in the target waveform.

To identify the transfer function, a set of step functions with different amplitudes is applied. Based on the response, the Matlab function `tfest' is used to find the linear dynamic response. The number of zeros and poles is varied to get a minimal system with good response. 

The system identification procedure is carried out semi-automatically, by running first the Arduino script \texttt{Identify.ino} and the Matlab script \texttt{nidaq\_i\-den\-tify.m} (to control the pump and read the resulting flow rates, respectively), and then the Matlab script \texttt{IdentifyPump.m} for the actual identification (see supplementary material for the code).

\subsubsection{Control}
\begin{figure}
\centering
	\includegraphics[width=0.75\textwidth]{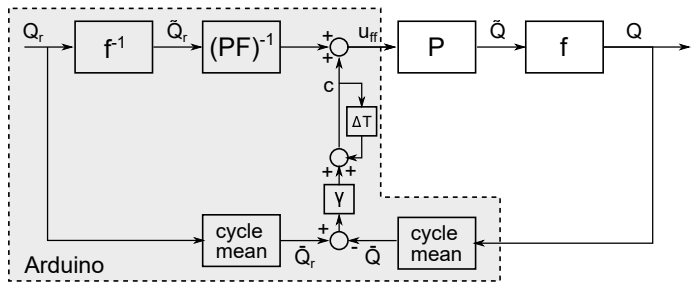}
	\caption{Control scheme of the system.}
	\label{fig:controlScheme}
\end{figure}

Figure~\ref{fig:controlScheme} shows the control scheme that was used. The control of the pump is based on a feed forward controller to achieve the desired tracking behaviour, and a run-to-run controller that adds a constant voltage to the feed forward signal during each period. The value of this constant, $c$, is adapted at the end of each period as $c := c + \gamma(\bar{Q}_\mathrm{r} - \bar{Q})$, where $\bar{Q}_\mathrm{r}$ is the mean reference flow, $\bar{Q}$ is the measured mean flow, and $\gamma$ is a proportionality constant regulating the strength of the run-to-run controller, set to 5.0 in our case (trial-and-error). 
%BK 
In the control scheme block $\Delta T$ is a delay of one pulse cycle.
The advantage of using a run-to-run controller is that it will result in a correct mean flow, even if there are estimation errors in the non-linear mapping and the dynamic response (equations \ref{eq:nonlinear} and \ref{eq:transferfunction}, respectively). The feed forward signal, $u_{\rm ff}$, is calculated as:
\begin{equation}
\label{eq:ffw}
	u_{\rm ff} = \left(P(s)F(s)\right)^{-1}\tilde{Q}_\mathrm{r} + c, \ F(s) = \frac{\omega_c^2}{(s + 2\zeta\omega_c s + \omega_c^2)(s/\omega_c+1)}
\end{equation}
The low pass filter $F(s)$ is needed to make the complete system strictly proper and hence invertible, $\omega_c$ is the cut off frequency of this filter in radians per second (set to 10~Hz in our case, using trial-and-error to balance between a sufficiently fast response on the one hand, and highly oscillatory behaviour on the other)
 
%% file: experiments.tex
\section{Experiments}
The experiments are divided into an identification part, and the actual waveform testing. Because the output flow rate of the gear pump in our set-up depends not only on the input voltage, but also on the resistance, inertia and compliance in the flow loop (e.g. flow phantom geometry, material, test fluid properties), the system identification must be done every time the flow loop is changed.
\subsection{Identification}
\subsubsection{Non-linearity}
\begin{figure}
\centering
	\psfragfig{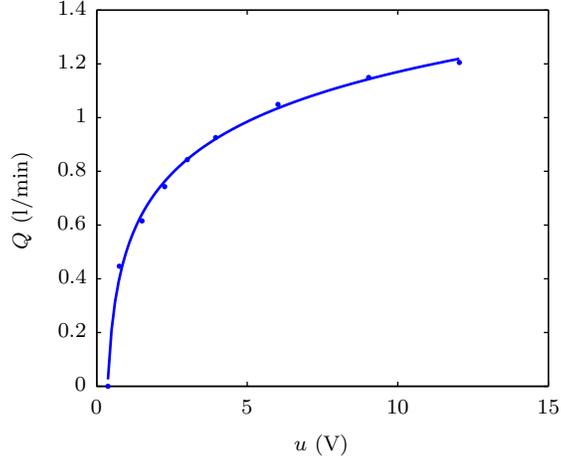}
	\caption{Relation between the input voltage and the measured flow in steady state (the dots represent measurement points, the drawn line shows the fitted function, with the weight vector as mentioned in the text).}
	\label{fig:NonLinFit}
\end{figure}

A set of driving voltages were set on the input of the pump system, and the resulting steady flow rates were measured. The result is shown in Figure~\ref{fig:NonLinFit}. The dots show the measured steady state values, and the line is the fitted function. The relation between the flow and the virtual flow is given by:
\begin{equation}
\label{eq:nonlinear}
Q = p_1\ln(p_2\tilde{Q}+p_3)+p_4.
\end{equation}
This relation is not based on the underlying physics, but on providing a good fit. The applied weight vector is $w = [0.5, 0.5, 1, 1, 5, 5, 1, 1, 1]$, these coefficients were chosen 
%BK by trial-and-error. 
to improve the fit in the working regime. 

\subsubsection{Linear system identification}
\begin{figure}
\centering
	%\psfragfig{DynFit_virt}
	\psfragfig{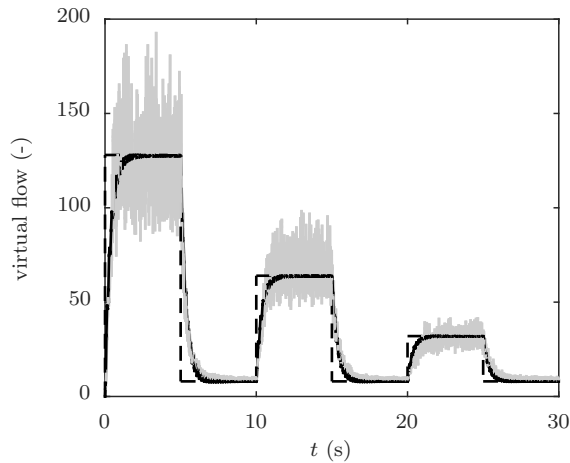}
	\caption{Response of the pump to step inputs of varying amplitude (virtual flow, in \textsc{pwm} units). Dashed line: input signal, grey: actual measurement data, black: fit of the linear system.}
	\label{fig:LinFit}
\end{figure}

A set of steps is applied to the pump and the actual flow is measured to identify the linear dynamic response. The non-linear mapping (\ref{eq:model}) is inverted to calculate the virtual flow in the step test. Figure~\ref{fig:LinFit} shows this virtual flow with the gray line. The black line is the fitted response. For the present fit a quantity like the root mean square error (RMSE) would not give much useful information, as its value would be mainly determined by the noise level of the measured signal. Analysing the signals in the frequency domain would be a better option, but is considered to be beyond the scope of this work. Therefore the quality of the fitted response was assessed visually, and only the final waveform was analysed quantitatively.
The corresponding transfer function is 
\begin{equation}
\label{eq:transferfunction}
 P(s) = \frac{4617}{(s+2.932)(s^2 + 2.6s + 1580)}.
\end{equation}
The first part of the denominator represents the exponential increase/decline due to a step, with a time constant of 2.9~sec, which can be seen in Figure~\ref{fig:LinFit}. The second part of the denominator coincides with the characteristic equation of a standard second order system, $s^2 + 2\zeta\omega_r + \omega_r^2$, with a relative damping of the characteristic equation, $\zeta = 0.03$, and a resonance frequency, $\omega_r = 6.3$ Hz (39.7~rad/s). The convolution of these two terms determines the behaviour of our linear system.
%BK
%
All the poles are in the left-half plane, which makes this linear transfer function stable. 

\subsubsection{Control}
The feed forward signal was calculated in Matlab using (\ref{eq:ffw}), and sent to the Arduino over a serial port (using the Matlab script \texttt{SendWaveform.m} and the Arduino script \texttt{Waveform.ino}, see supplementary material). The result is shown in Figure~\ref{fig:ffSig}. Note that roughly between $t = 0.4$s and $t = 0.5$s, the signal becomes negative. As mentioned in the previous section, our current setup cannot provide negative voltages, and these values will be cut at zero volt. As a result the signal is expected not to drop as quickly as needed, which would lead to a deviation of the mean flow if not corrected by the run-to-run controller. 
\begin{figure}
\centering
	\psfragfig{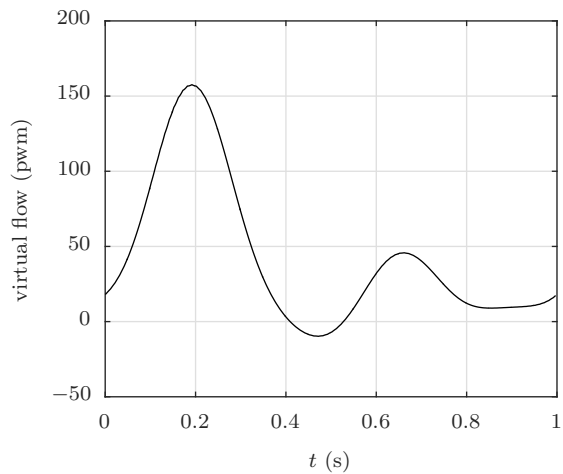}
	\caption{The feed forward signal in \textsc{pwm} units. Note: roughly between $t = 0.4$ and 0.5s the computed signal becomes negative, which is not possible to implement in the current system. It will be set to zero at these values.}
	\label{fig:ffSig}
\end{figure}

\subsection{Target waveform}
\begin{figure}
\centering
	\centering
	\psfragfig{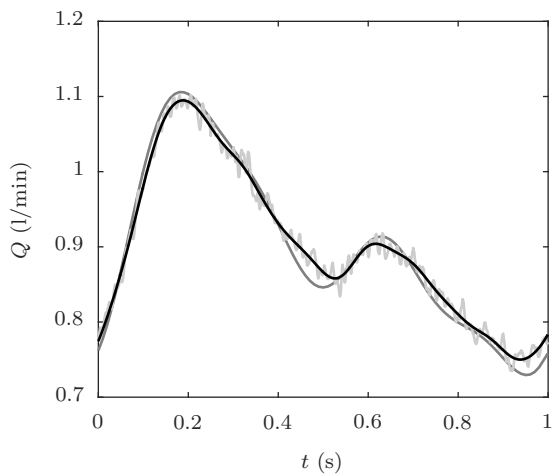}
	\caption{The measured waveform signal, low-pass filtered at 10~Hz (black line). The target waveform is shown in dark grey, the light grey line shows the raw measurement data.}
	\label{fig:MeasRefFilt}
\end{figure}

Figure \ref{fig:MeasRefFilt} shows the measured waveform signal, together with the target waveform. The flow rate was measured using the Transonic flowmeter system, and read out at a sampling frequency of $1\,000$~Hz. The filtering on the Transonic system was set to 160~Hz, and the signal was low-pass filtered in Matlab afterward, using a zero-phase filtering with a $5^\mathrm{th}$-order Butterworth filter with a cut-off frequency of 10~Hz.
The effect of the feed forward signal being restricted to positive values is clearly visible between $t = 0.4$s and $t = 0.5$s, but the overall performance is good. The origin of the deviation from the reference signal between 0.9 and 1.0~s (-3.5\% at $t = 0.97$~s) is unclear and should be investigated in more detail. The average error (average difference between the measured signal and the reference signal during one period) is 1.3\% of the target value, while the maximum error is 3.6\%, at $t = 0.47$s. This is within the accuracy of the Transonic flow sensor (4\% for the ME 6PXN in-line sensor).

%% file: conclusions.tex
\section{Conclusions and recommendations}
A programmable pulsatile flow pumping system was built, using an Arduino micro-controller to implement a feed forward signal computed in Matlab. Even though the used set-up is relatively simple, it was shown to be able to reproduce the desired waveform with an error of less than 3.6\% of the target value (1.3\% on average). This result shows that the assumption that the system can be modelled as a linear dynamic system with a static non-linear output function provides a good enough approximation of the true system for our application.

A drawback of the current system, and of systems of this type in general, is that the output flow rate not only depends on the input voltage, but also on the properties of the components present in the system, like the flow loop geometry and test fluid properties. This necessitates an identification step every time the geometry or test fluid are changed. Even though this identification step can be done semi-automatically (see code in supplementary material), and hence does not take much time, it should be done carefully, as the equation used to compute the virtual flow is exponential.

The observed resonance frequency at 6.3~Hz, and the inability of the current system to actively decelerate the flow (i.e.\ apply a negative voltage), limit the waveforms that can be reproduced with our set-up. The response of the system will decrease quickly at  frequencies higher than the resonance frequency. As a result, frequencies that are above 6.3 Hz in the reference signal need large driving voltages to be tracked correctly. The maximum voltage that can be applied to the DC motor in our set-up is 30~V, so in practice it will not be possible to track signals that exceed the resonance frequency.

Addition of a feedback controller is not expected to improve the tracking of the required waveform. The bandwidth of a feedback controller for the current transfer function cannot exceed approximately one third of the resonance frequency, as the phase margin would become too small and excessive oscillations would result. In the current situation that would mean a bandwith of approximately 2~Hz for feedback control, while the targetted waveform contains frequencies up to 4~Hz.

For the application in vascular access hemodynamics the limitation on the attainable waveforms does not pose a serious problem, as the frequencies encountered in the flow upstream of a vascular access site are typically below 5~Hz. For other applications it is important to keep in mind that not only the pump capacity, but also the dynamics of the system as a whole should be taken into account when selecting components.

To further improve the accuracy of the current system, iterative learning control (ILC) can be applied~\cite{bib:Arimoto1984}, taking advantage of the repetitive character of the targeted pulsatile flow. This goes beyond the concept of a simple scheme for the control of pulsatile flow, but it is expected to increase the performance of systems like the one described in this note. 
% BK
Also, changing the electronics from a one-quadrant setup to a four-quadrant setup would improve the performance. The flow could be actively stopped, while we are now limited by the damping present in the system. It would require more electronics to facilitate this.

For applications like the one discussed here, in which the targeted waveform contains a relatively large DC-component, also delivering this 
%BK
DC-
component separately can improve the accuracy. This can for example be done by combining two separate power supplies, or two separate pumps (see e.g. Tsai and Sava\c{s} \cite{ref:Tsai10}).

Finally, the availability of Matlab may be an issue in some situations. This should not be a problem, however, as the system identification, controller design and communication with Arduino and DAQ can alternatively be done with Python, which is freeware.